\documentclass[letterpaper]{article}

\usepackage{natbib,alifeconf}  
\usepackage{url,hyperref,cleveref}
\usepackage{booktabs}

\usepackage{lipsum}

\newcommand\blfootnote[1]{%
  \begingroup
  \renewcommand\thefootnote{}\footnote{#1}%
  \addtocounter{footnote}{-1}%
  \endgroup
}

\usepackage{xcolor} 
\usepackage{tabularx}
\usepackage{float}
\usepackage{csquotes}

\title{What Makes a Bacterial Model a Good Reservoir Computer? \\Predicting Performance from Separability and Similarity}

\author{
    Laura Alonso Bartolomé$^{1,2,3,4}$,
    Jean-Loup Faulon$^{1*}$, \and
    Xavier Hinaut$^{2,3,4*}$ \\
    \mbox{}\\
    $^1$University Paris Saclay, INRAE, AgroParisTech, MICALIS Institute, 78350 Jouy-en-Josas, France \\
    $^2$Centre Inria de l'Université de Bordeaux, France \\
    $^3$Institut des Maladies Neurodégénératives, Université de Bordeaux, CNRS, UMR 5293 \\
    $^4$LaBRI, Bordeaux INP, CNRS, UMR 5800, France \\
    $*$co-supervised the study;  
    laura.alonso-bartolome@inria.fr
} 

\begin{document}

\maketitle

\begin{abstract}
Biological systems are promising substrates for computation because they naturally process environmental information through complex internal dynamics. In this study, we investigate whether bacterial metabolic models can act as physical reservoirs and whether their computational performance can be predicted from dynamical properties linked to separability and similarity. We simulated the growth dynamics of five bacterial species, one yeast species, and 29 \textit{Escherichia coli} single-gene deletion mutants using dynamic flux balance analysis (dFBA), with glucose and xylose concentrations as inputs and growth curves as reservoir states. Computational performance was assessed on random nonlinear classification tasks using a linear readout, while reservoir properties linked to separability and similarity were characterised through kernel and generalisation ranks computed from growth-curve state matrices. Several microbial models achieved high classification accuracy, showing that bacterial metabolic dynamics can support nonlinear computation. Clear differences were observed between species, with some models converging more rapidly and others reaching higher maximum accuracy, revealing a trade-off between convergence speed and peak performance. In contrast, all \textit{E. coli} mutants were dominated by the wild-type model, suggesting that gene deletions reduce the dynamical richness required for efficient computation. The difference between kernel and generalisation ranks was generally associated with improved accuracy, but deviations across models and sensitivity at low rank values limited its predictive power in practice. Overall, these results show that bacterial metabolic models constitute promising substrates for reservoir computing and provide a first step towards identifying microbial strains with favourable computational properties for future experimental implementations.
\end{abstract}

Submission type: \textbf{Full Paper}\\

Code repository will be available after acceptance. 
Data/Code available at: \href{https://github.com/lauraalonsobart/alife2026-bacteria-reservoir-rank-difference}{github.com/lauraalonsobart/alife2026-bacteria-reservoir-rank-difference}
\blfootnote{\textcopyright  2026 [AUTHORS' NAMES]. Published under a Creative Commons Attribution 4.0 International (CC BY 4.0) license.}

\section{Introduction}

Biological systems are attractive substrates for computation because they naturally process environmental information through complex internal dynamics. In bacteria, this potential has long motivated efforts in synthetic biology to engineer cells capable of implementing specific information-processing behaviours \citep{Ahavi2026review}. Landmark studies demonstrated that synthetic genetic circuits could produce dynamical functions such as oscillations and bistability \citep{Elowitz2000oscillatory}, \citep{Gardner2000genetic-toggle}, and subsequent work has extended these approaches towards increasingly sophisticated forms of microbial computing \citep{Ahavi2026review}. However, such bottom-up strategies rely on genetic engineering, often requiring substantial optimisation for each new application and facing important limitations in scalability, including metabolic burden and lack of orthogonality \citep{Ahavi2026review}.

\begin{figure}[h]
    \centering
    \includegraphics[width=3in]{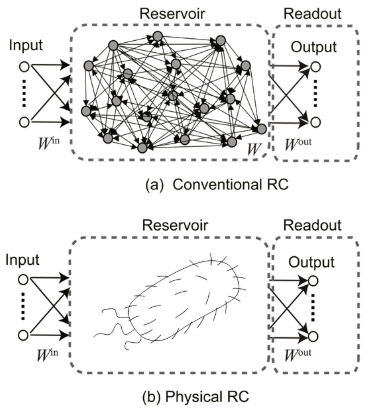}
    \caption{
        Conventional and physical reservoir computing frameworks. In conventional RC (a) the reservoir is a fixed weight system. In physical RC (b) the reservoir is a physical system with complex dynamics, for instance a bacterial population. Adapted from \cite{Tanaka2019review-physical-rc}.
    }
    \label{fig:conv-phys-rc}
\end{figure}

Physical reservoir computing (see \Cref{fig:conv-phys-rc}) offers an alternative approach in which computation emerges from the intrinsic dynamics of a system rather than from redesigning its internal architecture \citep{Tanaka2019review-physical-rc, Nakajima2020intro}. In the reservoir computing framework~\citep{lukovsevivcius2009reservoir}, inputs perturb a fixed dynamical system, called the reservoir, and only a readout layer is trained to map the resulting states to the desired output \citep{Tanaka2019review-physical-rc, Nakajima2020intro}. Because the reservoir itself does not need to be trained, this framework can exploit a wide variety of physical substrates with sufficiently rich nonlinear dynamics \citep{Tanaka2019review-physical-rc, Nakajima2020intro}. More broadly, reservoir computing has also been discussed from an evolutionary perspective, suggesting that natural systems may already possess properties compatible with information processing in this sense \citep{Seoane}.

In this context, bacterial populations are promising candidates for physical reservoir computing. Rather than modifying their regulatory, signaling or metabolic networks, one can directly exploit their natural dynamics as computational resources, as illustrated in \Cref{fig:conv-phys-rc}. A recent study by \cite{Ahavi2026bacterial-reservoir-preprint} demonstrated this idea experimentally by using nutrient inputs and growth dynamics of \textit{Escherichia coli} to perform classification and regression tasks. The same study also showed a practical sensing application, in which bacterial responses enabled discrimination between plasma samples from COVID-19 patients that later developed moderate or severe forms of the disease \citep{Ahavi2026bacterial-reservoir-preprint}.

These results raise a broader question: beyond \textit{E. coli}, what makes a microbial system suitable for reservoir computing? More specifically, can the computational performance of a bacterial strain be predicted from the properties of its internal dynamics? Related questions have previously been investigated for neural circuit models by Legenstein and Maass \citep{Legenstein2007edge-chaos-performance}, who showed that computational performance depends on the separability/similarity trade-off:
a balance between being able to distinguish meaningful differences without being disturbed by noise. Particularly effective regimes are supposed to occur near the ``edge of chaos'' \citep{Legenstein2007edge-chaos-performance}.

In this paper, we apply an analogous analysis to bacterial models to determine whether these dynamical properties can also predict computational capacity in bacterial reservoirs. Our aim is to characterise bacterial strains using metrics related to kernel rank (separability) and generalisation rank (similarity), and to assess whether these metrics are informative predictors of reservoir performance. In the long term, this could help identify optimal bacterial strains and support the development of new biological experimental setups.

\section{Methods}

\begin{figure}[h]
    \centering
    \includegraphics[width=3.3in]{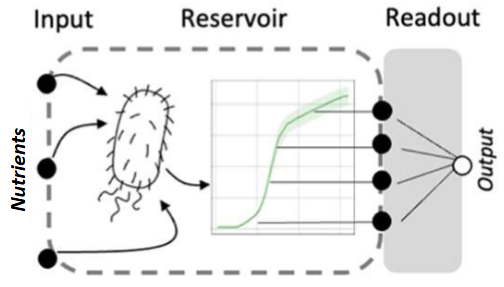}
    \caption{
        Bacterial reservoir setup. 
        Nutrients (e.g., sugars) are given directly as inputs to the bacterial population (model). Different growth curves were generated using dynamic flux balance analysis (dFBA) depending on these nutrients. At the end of the simulation, for each model, 200 time point values of the growth curve are fed to the readout to predict the output.
    }
    \label{fig:exp-meth-nut}
\end{figure}

The proposed implementation of bacterial reservoirs is inspired from \cite{Ahavi2026bacterial-reservoir-preprint}. As shown in \Cref{fig:exp-meth-nut}, nutrients are given as input and the growth dynamics of the bacterial population are interpreted as the reservoir state. In the present study, however, only computational models of the different bacterial species are tested. We used \emph{dynamic flux balance analysis} (dFBA) to predict growth curves, which act as reservoir states. We use such growth curves states to measure computational performance and to create the state matrices for the separability and similarity measures.
To produce the simplest possible experiment that could get significantly different responses from the bacteria we take only two different nutrients as inputs: glucose and xylose. There were found to have the highest effect on growth rate among the 28 metabolites tested in \cite{Ahavi2026bacterial-reservoir-preprint} and are known to produce what is known as diauxic growth. \emph{Diauxic growth} is a growth in two phases that is caused by the bacteria having different preference for the sugars\footnote{Various diauxic growth shifts depending on the type of sugar can be seen at \url{https://en.wikipedia.org/wiki/Diauxic_growth\#/media/File:Monod's_Diauxic_growth.gif}}. This shift when the bacteria change from consuming one sugar to the other can produce more complex dynamics and improve the amount of information we can get from the growth curve.

\subsection{Growth curve simulation: dFBA models}
To simulate the behavior of the bacteria and generate the growth curves from the nutrient concentration data we implemented a dFBA pipeline using COBRApy \citep{Ebrahim2013cobrapy}. FBA uses \emph{genome scale metabolic models} (GEMs) of organisms to simulate cell metabolisms and mathematically predict the fluxes of metabolites (including the increase in biomass) in steady state conditions under different constraints. Dynamic FBA involves simulating the evolution of the bacterial population over time by tracking the concentration of some molecules and iteratively running FBA simulations while modifying the constraints related to these molecules. In this study, we track the concentration of glucose and xylose. 

The growth curve simulation works as follows:
\begin{itemize}
  \item Uptake fluxes -- how fast a metabolite is being transported into the cells -- are calculated from the current concentration of glucose and xylose, and set in the model as lower bounds for the corresponding uptake reactions.
  \item FBA simulation returns growth rate and fluxes for glucose and xylose.
  \item These values are used to calculate the biomass growth in that time interval and the consumption of glucose and xylose. This allows to calculate the concentrations to be used in the next iteration.
\end{itemize}

The uptake fluxes for glucose and xylose were defined by the following kinetic equations: $$V_g=V_{g,max}\frac{G}{K_g+G}$$ $$V_x=V_{x,max}\frac{X}{K_x+X}\frac{1}{1+\frac{G}{K_{ig}}}$$

The transport of both sugars into the cells is assumed to follow Michaelis-Menten kinetics with an extra glucose dependent inhibitory term on xylose. The value for the related parameter $K_{ig}$ was selected to ensure no xylose consumption in the presence of glucose, as the later is known to produce strong catabolic inhibition of other sugars \citep{Kaplan2024glucose-xylose}. That is, when glucose is available, cells consume it first and only move onto consuming other sugars once glucose runs out. Additionally, the maximum oxygen uptake rate was limited to 5 $mmol/g_{_{DW}} \cdot h$  to reflect the low oxygen conditions cells would likely be subjected to in a plate reader which is where this kind of experiments are usually carried out. The parameters selected for this simulation and their sources can be found on Table~\ref{table:nut-uptake-param}.
Simulations were carried out for an experiment duration of 20 hours with a timestep $dt$ of 0.1 hours, resulting in 200 timepoints.

\begin{table}[ht]
    \centering
    \begin{tabular}{ccc}
        \toprule
        Parameter & Value & Source\\
        \midrule
         $V_{g,max}$ & 10.0 & \cite{Mahadevan2002dfba-coli}\\
         $K_g$ & 0.015 & \cite{Mahadevan2002dfba-coli}\\
         $V_{x,max}$ & 9.0 & \cite{Henson2014dfba-synth-micro}\\
         $K_x$ & 0.01 & \cite{Henson2014dfba-synth-micro}\\
         $K_{ig}$ & 0.01 & \cite{Kaplan2024glucose-xylose}\\
        \bottomrule
    \end{tabular}
    \caption{
        Nutrient uptake parameters for dFBA simulations.
    }
    \label{table:nut-uptake-param}
\end{table}

\subsection{Measuring computational performance}
\begin{figure}[h]
    \centering
    \includegraphics[width=3.3in]{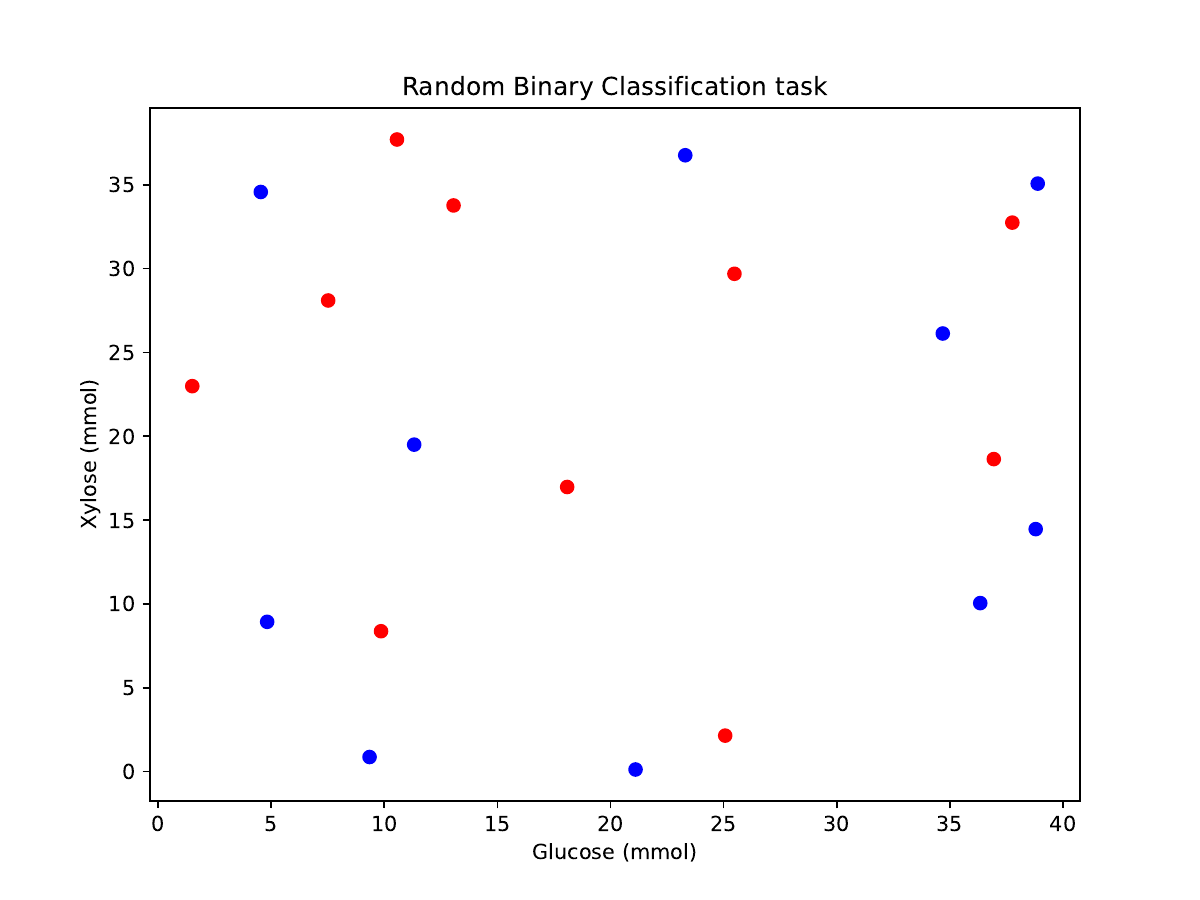}
    \caption{
        Example of a random binary classification task with 20 points.
    }
    \label{fig:classification-task}
\end{figure}

To act as a measure of computational performance the reservoir models were fed inputs consisting of different concentrations of two nutrients: glucose and xylose. Twenty combinations of concentrations were drawn from a uniform distribution of values between 0 and 45 mmol for both glucose and xylose. The bacterial models were fed noisy versions of these templates. Then, given the growth curves used as reservoir states, a linear readout was trained to perform 100 random classification tasks (i.e., 100 out of the $2^{20}$ possible binary classifications of the templates). Statistically, all 100 of these tasks are expected to be nonlinear in two dimensions, as random binary labeling of a 20-point dataset in a high-dimensional feature space almost surely results in a nonseparable decision boundary like in the example in \Cref{fig:classification-task}). To implement the readout we used the ridge classifier (\emph{RidgeClassifierCV}) from scikit-learn \citep{scikit-learn}, which optimize regularization parameter with cross-validation (alphas ranging from $10^{-3}$ to $1$).

\begin{figure}[h]
    \centering
    \includegraphics[width=3.3in]{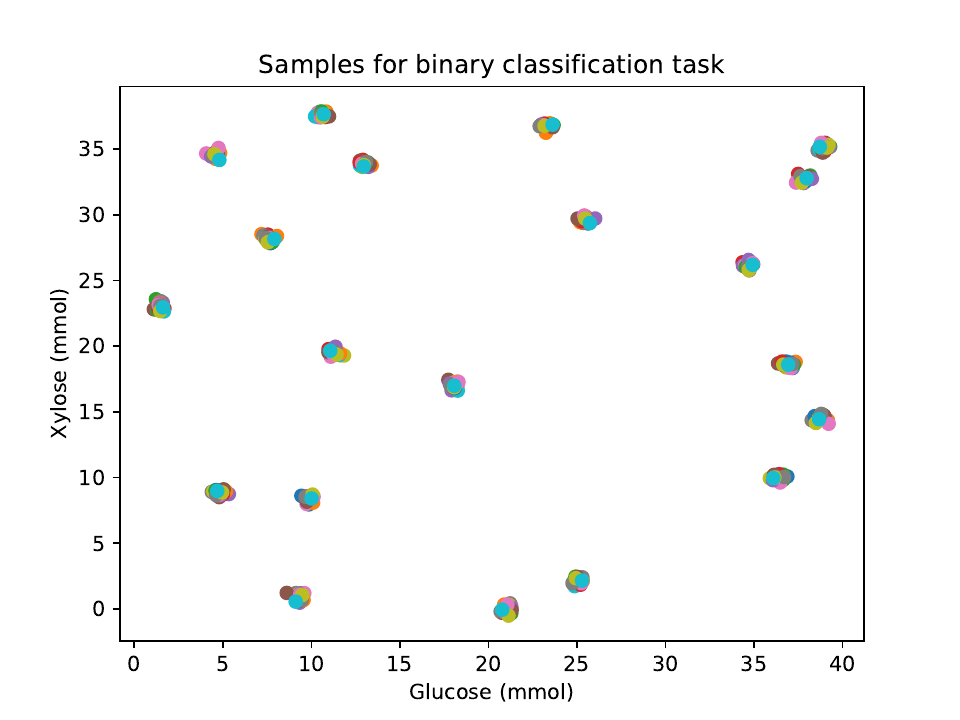}
    \caption{
        Input samples used to measure computational performance. The samples are aggregated in small clusters comprising 20 noisy inputs around their corresponding templates. This level of noise mimics experimental preparation error.
    }
    \label{fig:samples}
\end{figure}

In \Cref{fig:samples}, we see a representation of the noisy input versions of the templates.
The added noise was modelled as a small Gaussian perturbation of the concentration levels (mean 0, std 0.2), representing experimental error that may occur during the preparation of media with defined nutrient concentrations \citep{iec1995guide}. Mean accuracy for 100 such tasks was reported for each model from a set of 400 training samples and 100 test samples.

\subsection{Kernel and generalisation ranks}

We assessed the computational properties of each bacterial model using two rank-based metrics derived from the framework introduced by \cite{Legenstein2007edge-chaos-performance}: 
\emph{kernel rank}\footnote{Legenstein and Maass \citep{Legenstein2007edge-chaos-performance} use the term \emph{kernel quality}. We instead use \emph{kernel rank} to emphasise that the measure is defined as the rank of a state matrix and to simplify later discussion of rank differences.},
which reflects separability, and 
\emph{generalisation rank}\footnote{Legenstein and Maass \citep{Legenstein2007edge-chaos-performance} use the term \emph{generalization capability}. We instead use \emph{generalisation rank} to avoid confusion with generalisation in the machine learning sense of predictive performance on unseen data.},
which reflects similarity.
More specifically, kernel rank reflects the ability of the system to produce linearly separable states, whereas generalisation rank reflects the ability to produce similar states in response to noisy variations of the same input.
Accordingly, lower generalisation rank values indicate greater similarity among the corresponding states, whereas higher values indicate lower similarity.

For each input condition, the bacterial model generated a growth-curve time series. For a given input ensemble, these response trajectories were assembled into a state matrix, with one row per input, and columns corresponding to the sampled time points of the growth curve. The rank of this matrix was then used to quantify the diversity of the reservoir responses.

To estimate separability, 100 templates were randomly drawn from the previously mentioned uniform concentration distributions and applied to the bacterial models. The kernel rank was computed as the rank of the resulting state matrix. A high kernel rank indicates that distinct inputs are mapped to distinct, linearly independent response trajectories, which corresponds to strong separability.
To estimate similarity, 5 templates were selected and 20 noisy variants were generated from each template, yielding 100 input samples in total. The generalisation rank was computed from the corresponding state matrix. In this case, lower rank values indicate that similar inputs produce similar response trajectories, reflecting stronger generalisation\footnote{To understand the link to the VC-dimension, one can refer to \cite{Legenstein2007edge-chaos-performance} and this interactive tutorial from \cite{lauer2014vcdimhyperplane}:  \url{https://mlweb.loria.fr/book/en/VCdimhyperplane.html}}.
These two measures were used jointly to characterise the balance between discriminating meaningful differences in the input space and maintaining robustness to small perturbations.

\subsection{Metabolic models}

The genome scale metabolic models (GEMs) used in this study include five different bacterial species (\textit{Escherichia coli}, \textit{Bacillus subtilis}, \textit{Staphylococcus aureus}, \textit{Salmonella} pan-reactome and \textit{Yersinia pestis}) and one yeast (\textit{Saccharomyces cerevisiae}). Even though \textit{S. cerevisiae} is not a bacterial species we decided to add it to the study because it is also a unicellular organism that is very commonly used in microbiology labs and it is interesting to see if there are notable differences in eukaryotes. The specific strains and model IDs can be found in \cref{table:gem-list} in the appendix. It is important to note that the same uptake reactions were used for all models, even though the parameters were extracted from scientific publications concerning \textit{E. coli}. This is because glucose and xylose transport parameters are not available for all species, and different uptake kinetics might interfere when drawing comparisons.

Additionally, 29 variants of the \textit{E. coli} GEM that reflect mutations relevant to the experimental conditions were also tested. These mutants were identified using the single gene deletion method from COBRApy, which enabled the computation of growth rates for all possible single gene deletions. To focus on biologically meaningful phenotypes, only deletions that resulted in growth rates between 0.05 and 0.95 times that of the wild-type model were retained. Deletions with no significant effect on growth or those that were excessively deleterious (leading to near-zero growth) were excluded.

\subsection{Observing evolution over time}
The default time for the simulations was 20 hours so that there was enough time to capture the full growth dynamics of most models even if some had slower growth. With a timestep of 0.1 hours, this results in 200 time points which constitute the state of the reservoir for each input. This state can then be used to solve the tasks and measure the accuracy of the model, or compute the kernel and generalisation ranks. 
However, while we evaluated the accuracy achieved by the models over a 20-hour experiment, it is also important to determine whether comparable accuracy can be obtained with a shorter experiment. To investigate this, we recalculated all metrics using progressively increasing numbers of time points, up to the original 200, allowing us to track how model accuracy and the ranks evolved over time.

\section{Results}

\begin{figure}[h]
    \centering
    \includegraphics[width=3.3in]{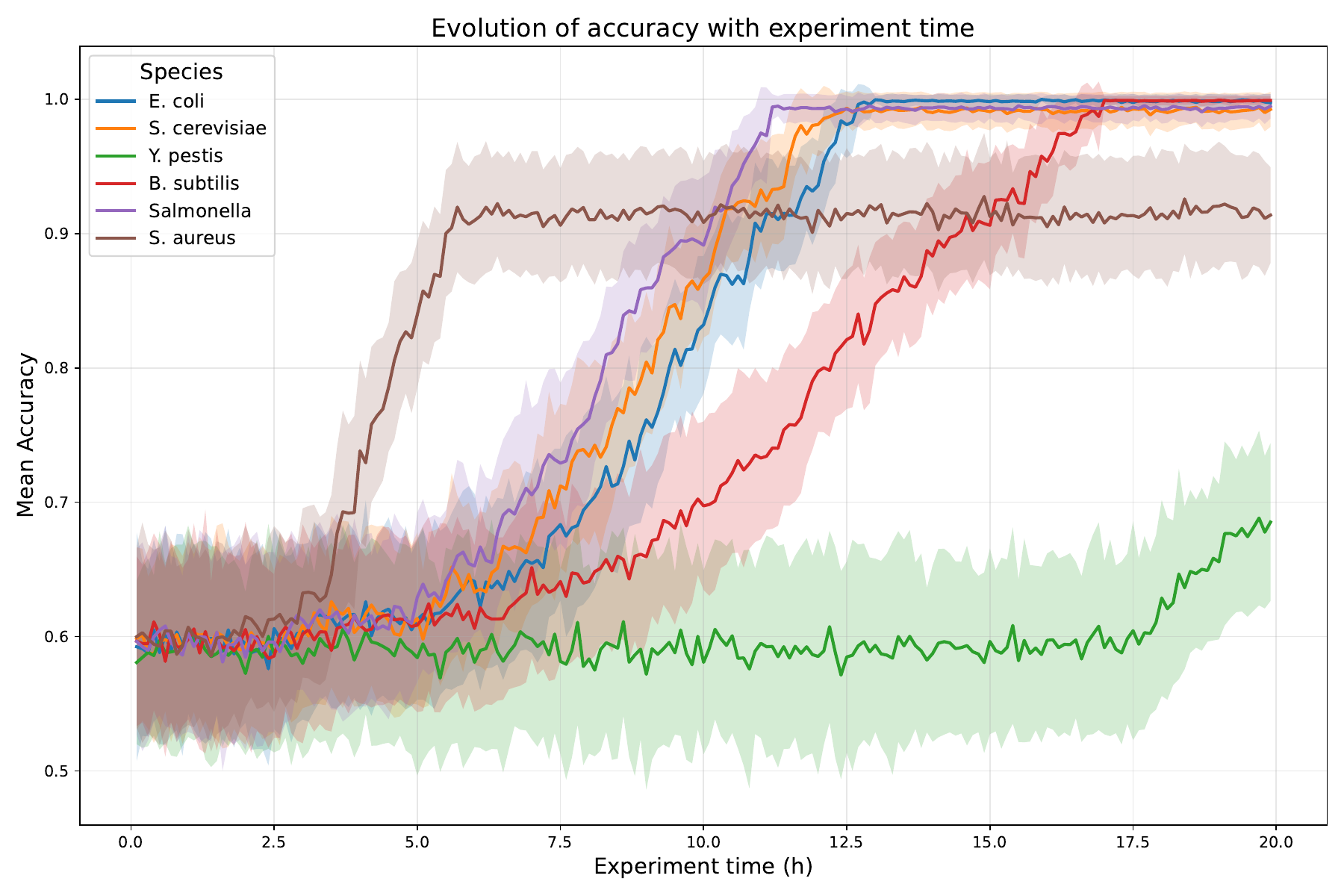}
    \caption{
        Accuracy over time for the different microbial species and one yeast species. Shaded areas represent the standard deviation around each curve. The x-axis spans 20 h, with one time point every 0.1 h.
    }
    \label{fig:evo-acc-diff-species}
\end{figure}

In \Cref{fig:evo-acc-diff-species}, we can see the evolution of mean accuracies for the classification tasks of the different microbial models. All species start at a baseline level around 0.6 of accuracy: this matches the accuracy obtained when using only the input without the bacteria model: i.e.,
the two sugar concentration values are directly fed to the linear readout. As the experiment time increases and growth dynamics are unveiled, the linear readout is able to produce better accuracies for these non-linear classification tasks with some species reaching almost perfect accuracy. This demonstrates the ability of the bacterial models to perform non-linear computations.

Visualizing the evolution over time allows us to see differences between the four models that reach near perfect accuracy in the 20 hour experiment. While \textit{Salmonella} and \textit{B. subtilis} perform similarly in longer experiments, \textit{Salmonella} is able to reach this level around five hours earlier, which is preferable if minimal time is the main objective. We also see that even though the \textit{S. aureus} model has the fastest growing accuracy it reaches a plateau at approximately 0.92 and is unable to improve further. Lastly, \textit{Y. pestis} is the most underperforming of the evaluated species, only starting to improve accuracy from the baseline level at around 17.5 hours.

\begin{figure}[h]
    \centering
    \includegraphics[width=3.3in]{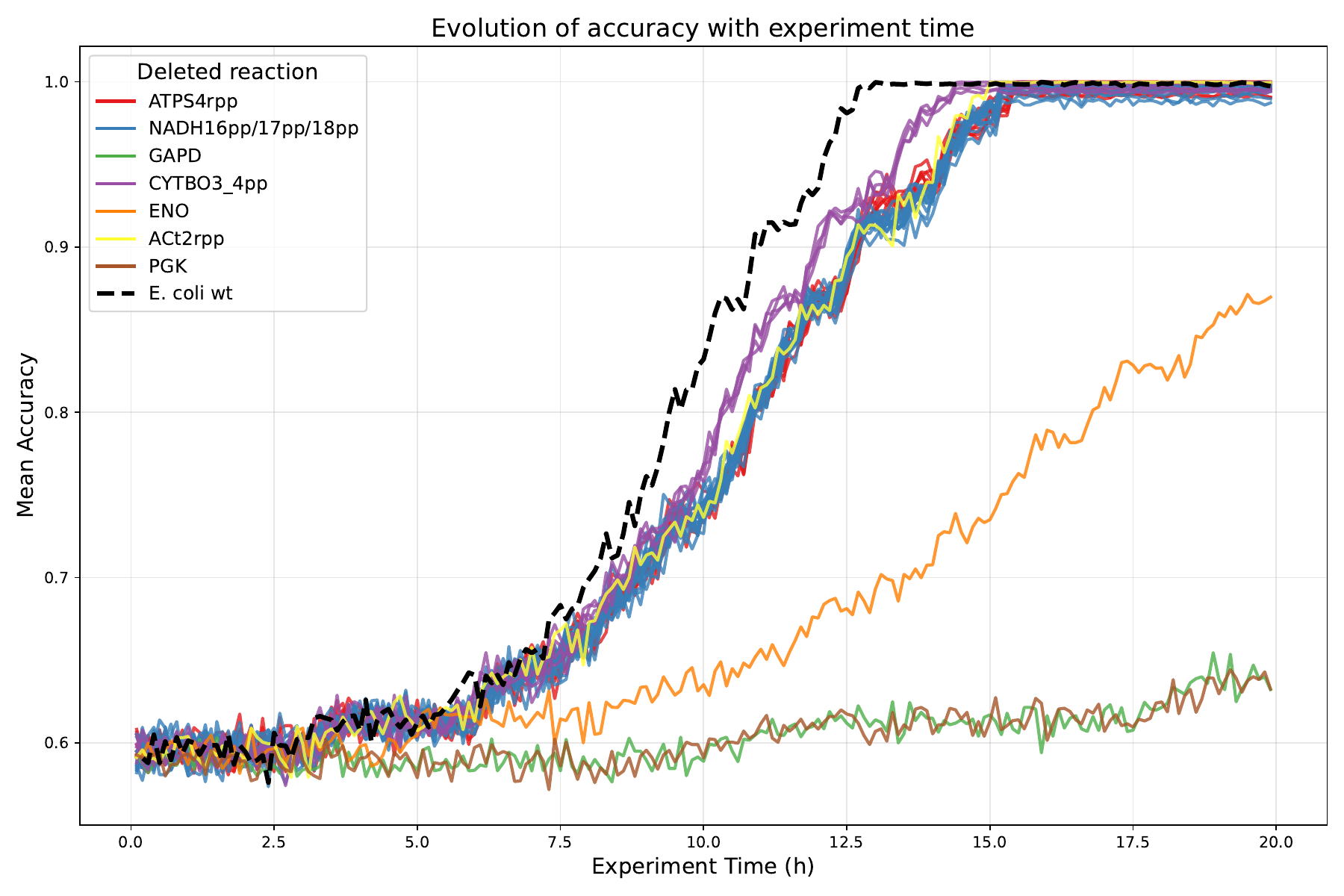}
    \caption{
        Evolution of accuracy with experiment time for the \textit{E. coli} mutants. The color of a mutant corresponds to the reaction that is affected by the gene deletion. The legend includes the BiGG ID of the affected reactions which comprise several mutants in the case of big enzymatic complexes. The data for the wild type strain is represented by the black dashed line. Standard deviations are not shown for visual clarity, a version with standard deviations can be found in the appendix (\Cref{fig:evo-accuracy-sd-mutants-e-coli}).
    }
    \label{fig:evo-accuracy-mutants-e-coli}
\end{figure}

Out of the 29 \textit{E. coli} mutants that were evaluated, we only get a handful of different behaviours (\Cref{fig:evo-accuracy-mutants-e-coli}). Firstly because some of these genes encoded proteins that are part of bigger enzymatic complexes like NADH dehydrogenase, ATP synthase and cytochrome oxidase so the elimination of either of them results in the same functional consequence which is the elimination of those reactions. In the figure, mutants that affect the same reaction are represented by the same color. Secondly, because some others encode enzymes that are involved in the same pathway, like Glyceraldehyde 3-phosphate Dehydrogenase (GAPD) and Phosphoglycerate Kinase (PGK) being consecutive steps of the more common Embden–Meyerhof–Parnas glycolysis pathway. The absence of either of these enzymes diverts the flux of sugar degradation to the Entner-Doudoroff pathway producing similar results.

None of the mutants produce better results than the wild type strain but some are clearly more damaging to the models ability to produce dynamics that allow accurate classification of the inputs. The worse performing mutants (PGK, GAPD and, to a lesser extent, ENO) are those that affect the glycolysis pathway which integrates the input sugars into the broader metabolic network. Mutants that affect enzymes involved in other pathways like the currency metabolites ATP and NADH can slow down growth but seem to not so critically affect the richness of the internal dynamics.

\subsection{Separability and similarity}
We then analysed the temporal evolution of the separability and similarity properties. As shown in Figure 6, the kernel rank, the generalisation rank, and their difference—which we term the rank difference—change over time. Consistent with expectations, the kernel rank rises more rapidly than the generalisation rank, while accuracy increases as the difference between the two becomes larger.

\begin{figure}[h]
    \centering
    \includegraphics[width=3.3in]{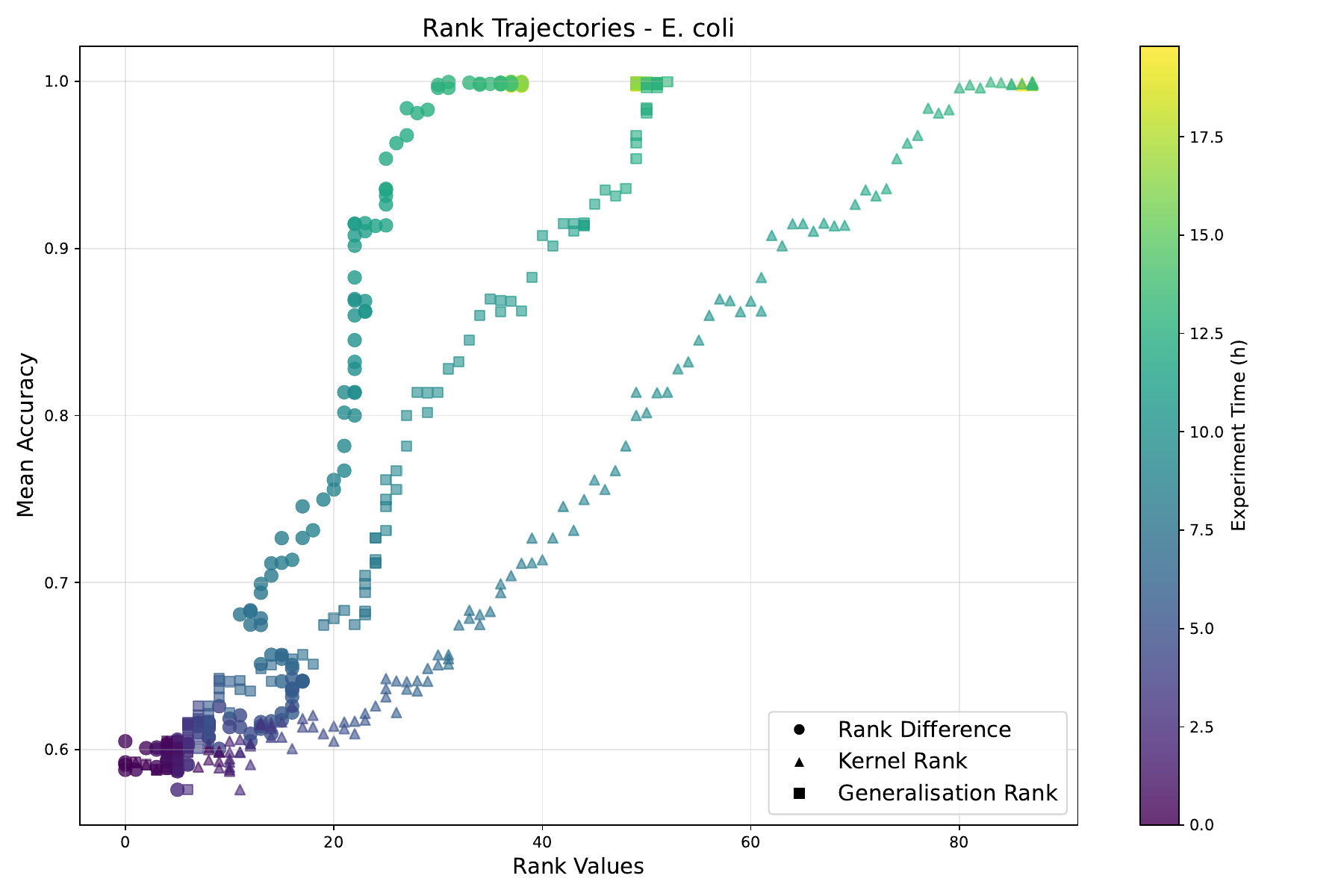}
    \caption{
        Trajectories of kernel and generalisation ranks individually next to their difference. Color gradient indicates time.
    }
    \label{fig:trajectories-kernel-gen}
\end{figure}

\begin{figure}[h]
    \centering
    \includegraphics[width=3.3in]{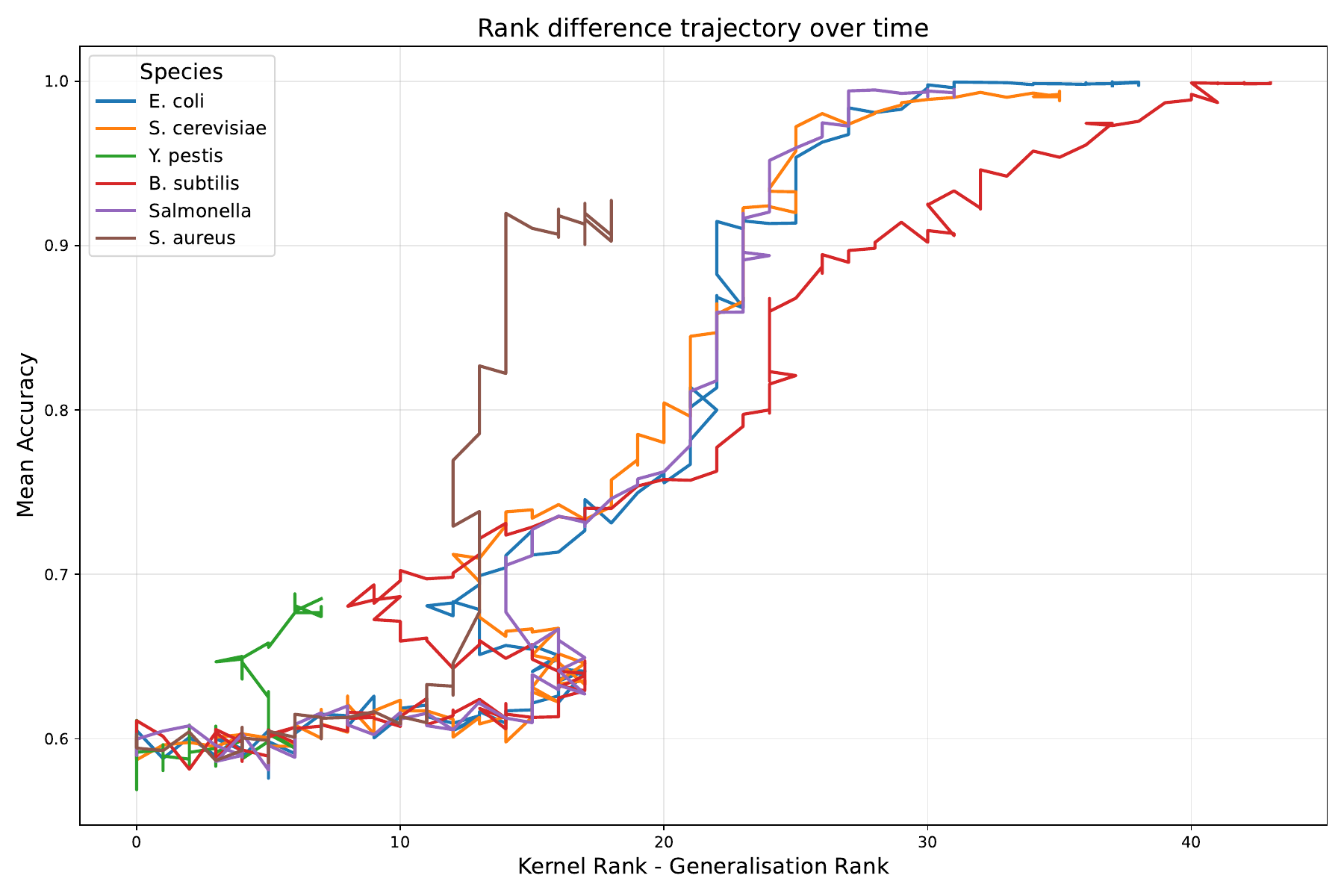}
    \caption{
        Relationship between rank difference and mean accuracy for classification tasks in the different microbial models.
    }
    \label{fig:rank-diff-acc}
\end{figure}

\Cref{fig:rank-diff-acc} shows the trajectory of the rank difference for the different microbial species. Two species stand out. \textit{S. aureus} reaches relatively high accuracies at low rank differences, with a sharp increase in accuracy occurring as the rank difference rises only slightly from 12 to 14. By contrast, \textit{Y. pestis} appears unable to generate large rank differences, although its slight improvement in accuracy at the end of the experiment is still achieved at very low rank difference values. The other models follow broadly similar trajectories, even if some take longer than others to reach higher accuracies. \textit{B. subtilis} deviates slightly, requiring a somewhat larger rank difference to attain accuracy levels comparable to those of the other models.

\begin{figure}[h]
    \centering
    \includegraphics[width=3.3in]{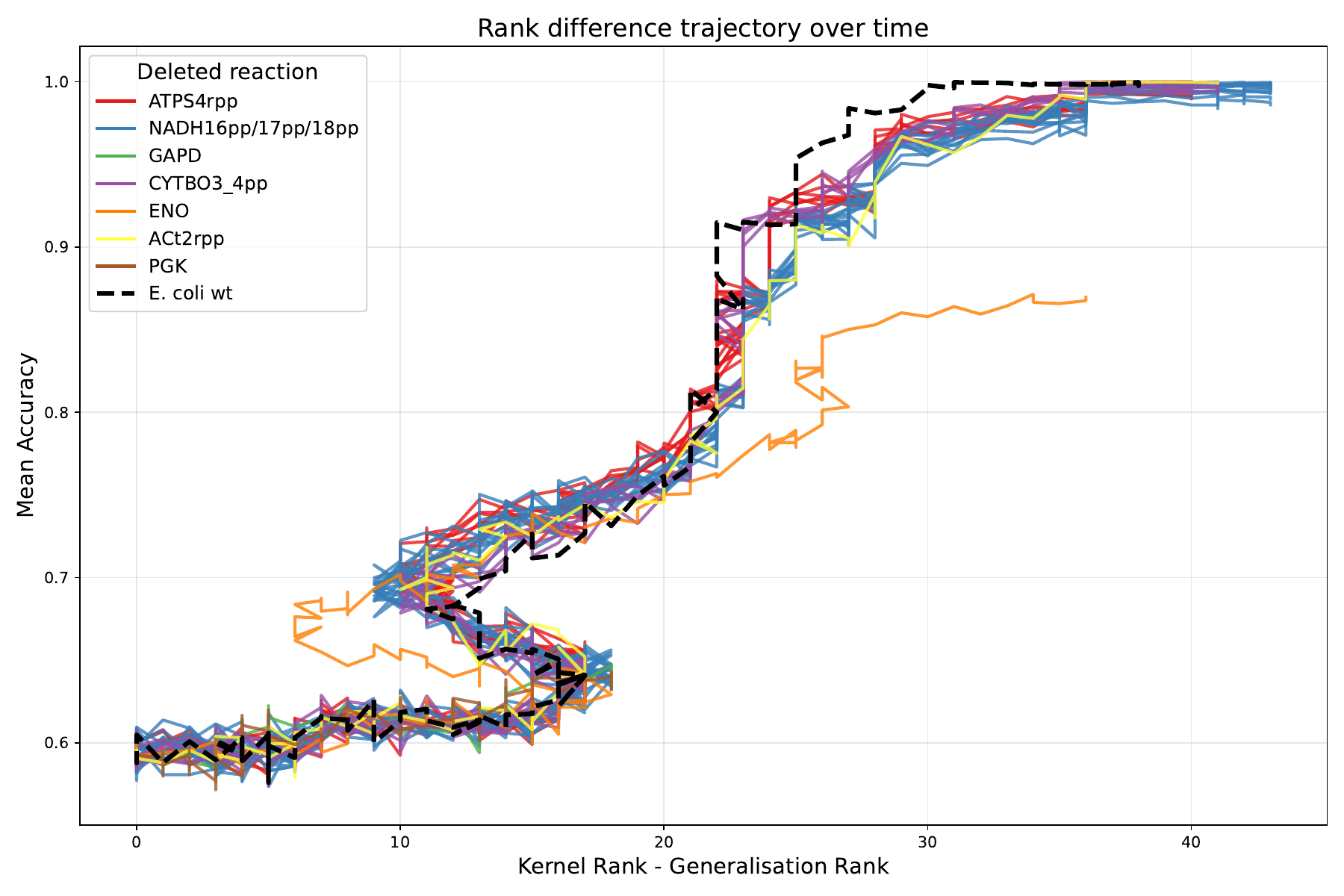}
    \caption{
        Relationship between rank difference and performance in the \textit{E. coli} mutants. The data for the wild type strain is represented by the black dashed line.
    }
    \label{fig:rank-diff-perf-ecoli-mutants}
\end{figure}

Lastly, \Cref{fig:rank-diff-perf-ecoli-mutants} shows the trajectories of the \textit{E. coli} mutants. In this case we see a clear aggregation of most models on a similar trajectory. The only exception is the enolase mutant which underperforms at equal rank differences when compared to the other models. Conversely, while GAPD (brown) and PGK (green) never reached higher rank differences or accuracies they do not deviate from the general trajectory. 

Over time, the rank difference exhibits an S-shaped pattern, already apparent in the previous figure and even clearer here. This temporary reversal in the trend arises from the diauxic shifts of two templates, and their corresponding 40 samples, used for the generalisation rank calculation, occurring in close succession.
Because this corresponds to a dynamically informative regime, the generalisation rank temporarily rises more rapidly, reducing the rank difference. A similar effect occurs at later time points, but it is only faintly visible in the ENO mutant. This suggests that the measure is especially sensitive at low rank values and may therefore be less reliable for predicting accuracy in that regime.

More generally, increases in rank difference appear to correlate with higher accuracy. However, while many models follow a similar trajectory, deviations in some cases complicate the use of this measure for model comparison. Its sensitivity to the specific samples used for rank computation also limits its usefulness at low rank values.

\section{Discussion and future work}

As some recent work has shown~\cite{Ahavi2026bacterial-reservoir-preprint}, bacteria have the  potential to be used as physical reservoirs. The main aim of this study was to explore the computational capability of different species and mutated strains and to characterise properties commonly measured in reservoir computing systems. Additionally, we wanted to test whether these separability and similarity properties were good predictors of a model's performance.

First, we found that many of these microbial models are capable of supporting nonlinear computations and achieve strong performance on classification tasks with highly nonlinear structure.
The results showed that some species performed better in terms of how quickly they reached higher accuracies, whereas others achieved higher maximum accuracies (see Figure~\ref{fig:evo-acc-diff-species}).

Secondly, the existence of these different behaviours allows selecting a model depending on whether faster convergence or higher maximum accuracy is considered more valuable. 
This indicates a trade-off between convergence speed and peak performance, allowing model selection based on which objective is considered more important. 
In multi-objective optimization terms, the Pareto front corresponds to the set of non-dominated models, that is, those for which improving convergence speed would come at the expense of maximum accuracy, and vice versa.
From the fastest to the max accuracy, the three strains on the Pareto front are : S. aureus, Salmonella and E. coli. 
In contrast, as shown in Figure~\ref{fig:evo-accuracy-mutants-e-coli}, all \textit{E. coli} mutants appear to be dominated by the wild-type model. This is consistent with the idea that the ability to generate a wide range of responses to different inputs is an important property of living systems. It is therefore not surprising that mutants carrying gene deletions perform worse than the wild type. In other words, a greater information-processing capacity may be advantageous in changing environments \citep{Seoane}. This may explain why the wild species display a clearer Pareto front, as is commonly observed in evolution when different traits are optimised at the expense of one another.

Lastly, while an increase in the difference between the kernel and generalisation ranks was found to be closely related to an increase in accuracy, deviations in some models and sensitivity to specific samples in low rank values diminish the potential of this metric as a predictor of accuracy in practice.

This opens up future work directions in areas like network analysis where we could investigate if any measurable properties of the metabolic networks like degree distribution or centrality are good predictors of the strains' computational capabilities. Additionally, other measures typical of reservoir systems like Information Processing Capacity (IPC) were not explored in the present study because they can only be applied to setups that handle sequential inputs. 
Future work will also involve developing a bacterial reservoir computing implementation that can process sequential data.

\section{Appendix}

Below we include a table with the models used in the study and reproductions of the figures with standard deviations for the cases where they were not included for visual clarity in the main text.

\begin{figure}[H]
    \centering
    \includegraphics[width=3.3in]{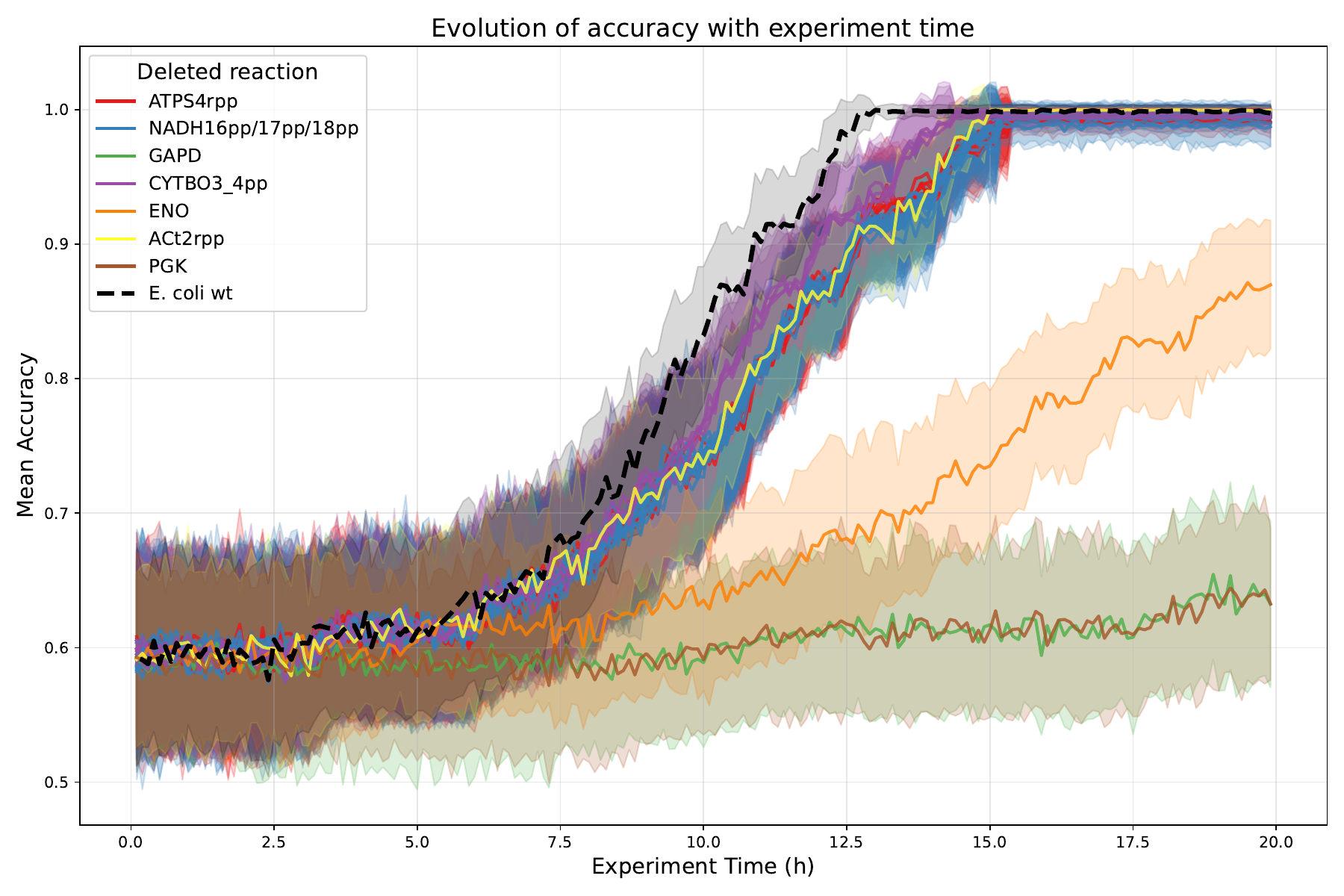}
    \caption{
        Evolution of accuracy with experiment time in the \textit{E. coli} mutants. The data for the wild type strain is represented by the black dashed line. The shaded areas represent standard deviation.
    }
    \label{fig:evo-accuracy-sd-mutants-e-coli}
\end{figure}

\begin{figure}[H]
    \centering
    \includegraphics[width=3.3in]{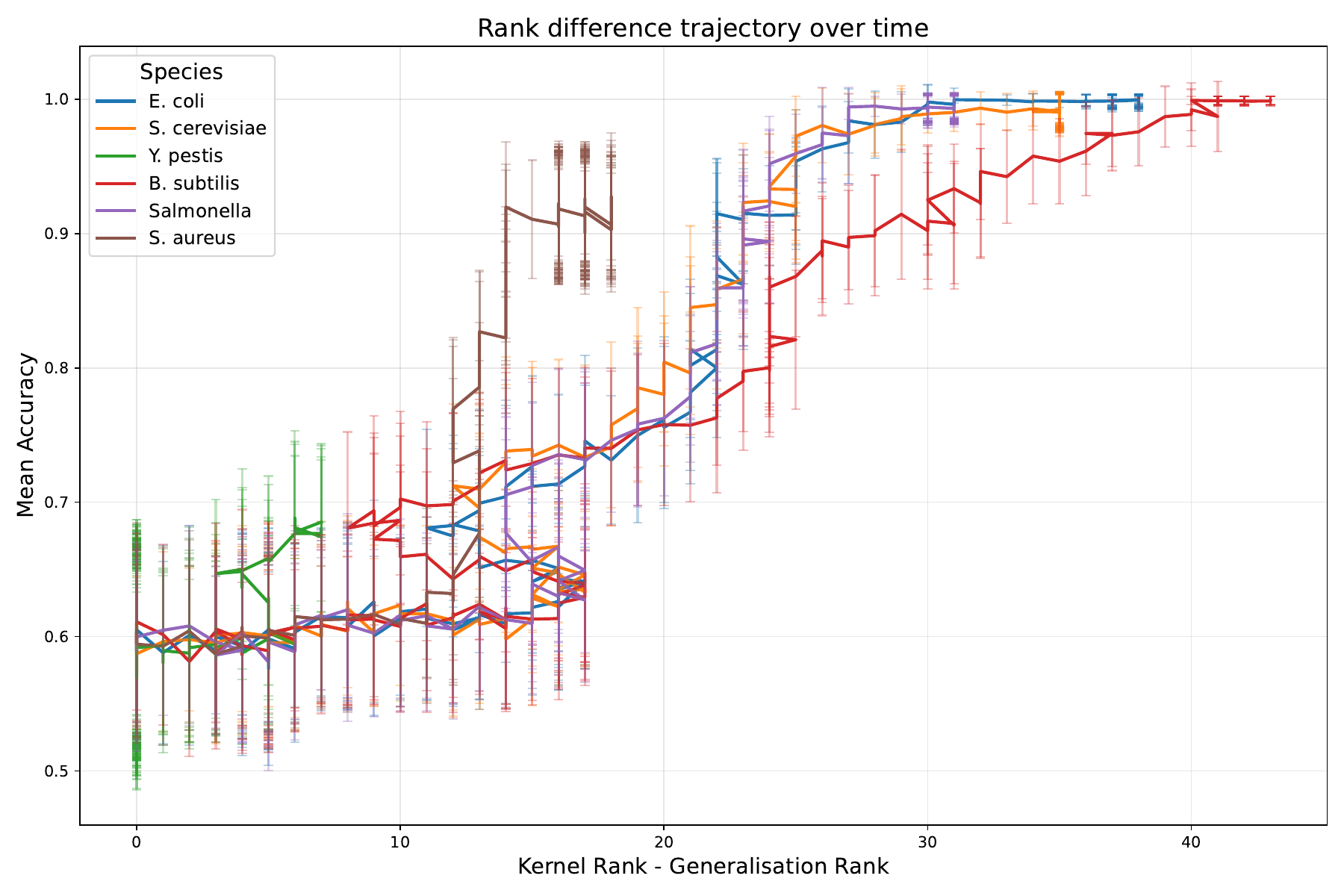}
    \caption{
        Relationship between rank difference and performance in the different species. Error bars represent standard deviation.
    }
    \label{fig:rank-diff-sd-perf-species}
\end{figure}

\begin{figure}[H]
    \centering
    \includegraphics[width=3.3in]{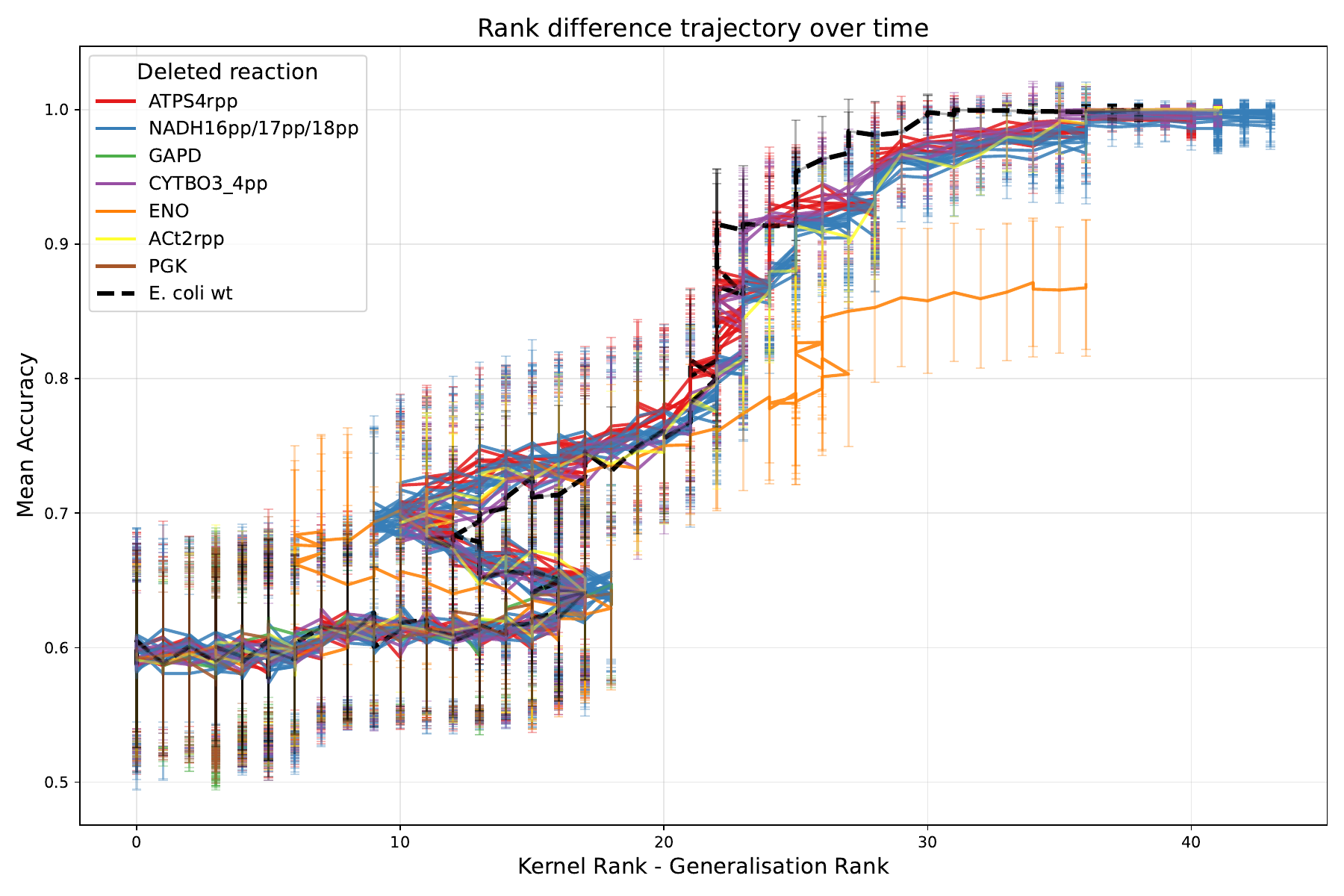}
    \caption{
        Relationship between rank difference and performance in the \textit{E. coli} mutants. The data for the wild type strain is represented by the black dashed line. Error bars represent standard deviation.
    }
    \label{fig:rank-diff-perf-sd-ecoli-mutants}
\end{figure}

\begin{table}[h]
    \centering
    \setlength{\tabcolsep}{4pt} 
    \renewcommand{\arraystretch}{1.2} 
    \begin{tabularx}{\columnwidth}{>{\raggedright\arraybackslash}X>{\raggedright\arraybackslash}X >{\raggedright\arraybackslash}X}
        \toprule
        BiGG ID & Short Name & Organism \\
        \midrule
        iML1515 & E. coli & E. coli str. K-12 substr. MG1655 \\
        iMM904 & S. cerevisiae & Saccharomyces cerevisiae S288C \\
        iPC815 & Y. pestis & Yersinia pestis CO92 \\
        iYO844 & B. subtilis & Bacillus subtilis subsp. subtilis str. 168 \\
        iYS1720 & Salmonella & Salmonella pan-reactome \\
        iYS854 & S. aureus & Staphylococcus aureus subsp. aureus USA300\_TCH1516 \\
        \bottomrule
    \end{tabularx}
    \caption{
        Genome-scale metabolic models (GEMs) used in the study.
    }
    \label{table:gem-list}
\end{table}

\section{Acknowledgements}
We would like to thank Thierry Vieville for early discussions on VC-dimension and related measures. We also acknowledge the following fundings: INRAE-Inria PhD grant, PEPR B-BEST France 2030 program (grant number ANR-22-PEBB-0008), the EU HORIZON BIOS program (grant number 101070281) and the DeepPool project (ANR-21-CE23-0009).

\footnotesize
\bibliographystyle{apalike}
\bibliography{bibliography}

\end{document}